# Measuring Bulk Flows in Large Scale Surveys

**Hume A. Feldman and Richard Watkins**
*Physics Department, University of Michigan, Ann Arbor, MI 48109, USA*

## ABSTRACT

We follow a formalism presented by Kaiser [1] to calculate bulk flows in large scale surveys. We apply the formalism to a mock survey of Abell clusters á la Lauer & Postman [2] and find the bulk velocities in a universe with CDM, MDM and IRAS–QDOT power spectra. We calculate the velocity variance as a function of the 1–D velocity dispersion of the clusters and the size of the survey.

## 1. Introduction

Recently Lauer and Postman (LP) [2] presented an analysis where they measured the reflex motion of the local group with respect to the 15000 km/s Abell Cluster frame. The velocity inferred is a $3\sigma$ result compared to the local group velocity in the CMB frame. They conclude that if the CMB dipole is doppler in origin one can infer the bulk velocity of the Abell Cluster frame in the cosmic frame to be $\approx 730$ km/s. They made a peculiar velocity survey using the brightest cluster galaxy (BCG) of 120 Abel clusters as distance indicators with peculiar velocity dispersion of 16% of the radial redshift distance. Their survey is volume limited and complete to all clusters with an elliptical BCG in a sphere of 15000 km/s radius. They found that the $L - \alpha$ relation showed a dipole in the sky directed $\approx 80°$ from the direction of the CMB dipole and contend that this velocity of the Abell cluster frame relative to the cosmic rest frame (as defined by the CMBR) represents the true velocity on these scales. Given present theories it is quite surprising to find as large bulk velocities on such scales. Strauss [3] presents Monte–Carlo simulations of peculiar velocity data to find the probability that different theoretical models can explain the data, we approach the problem semi–analytically to test various cosmological models and survey results.

We present an analysis based on work done by Kaiser [1] to infer what amplitude bulk flows one should measure given a specific velocity power spectrum. We show results for power spectra from the IRAS-QDOT survey [4], the BBKS CDM model [5] and from simulations of mixed dark matter (MDM) [6]. We show that the velocities are quadrature superposition of the noise that is inversely proportional to the number of data points and a convolution of the velocity power spectrum and the window function which is determined by the geometry of the survey. Clearly, the more power there is on large scales, the larger the expected velocities one would observe, therefore we chose as one of the power spectra we analyze the IRAS–QDOT plus the $1\sigma$ error bars.

In §2 we present the analysis and calculate the window function for an ensemble of observers surveying a similar sample to the LP survey and present the power spectra we used. In §3 we discuss the large $N$ limit analytically. In §4 we present the results for a LP type survey, extend the analysis to similar surveys of different sizes and accuracy and compare with analytical results. We conclude in §5.

## 2. Analysis

We would like to calculate what one would expect to measure for the net streaming velocity of a sample of clusters given a universe with a velocity power spectrum $P_v(k)$. To this end, we assume an ensemble of observers each of whom measures the line-of-sight velocities of each cluster in a "mock" cluster sample constructed to mimic the features of the actual sample observed.

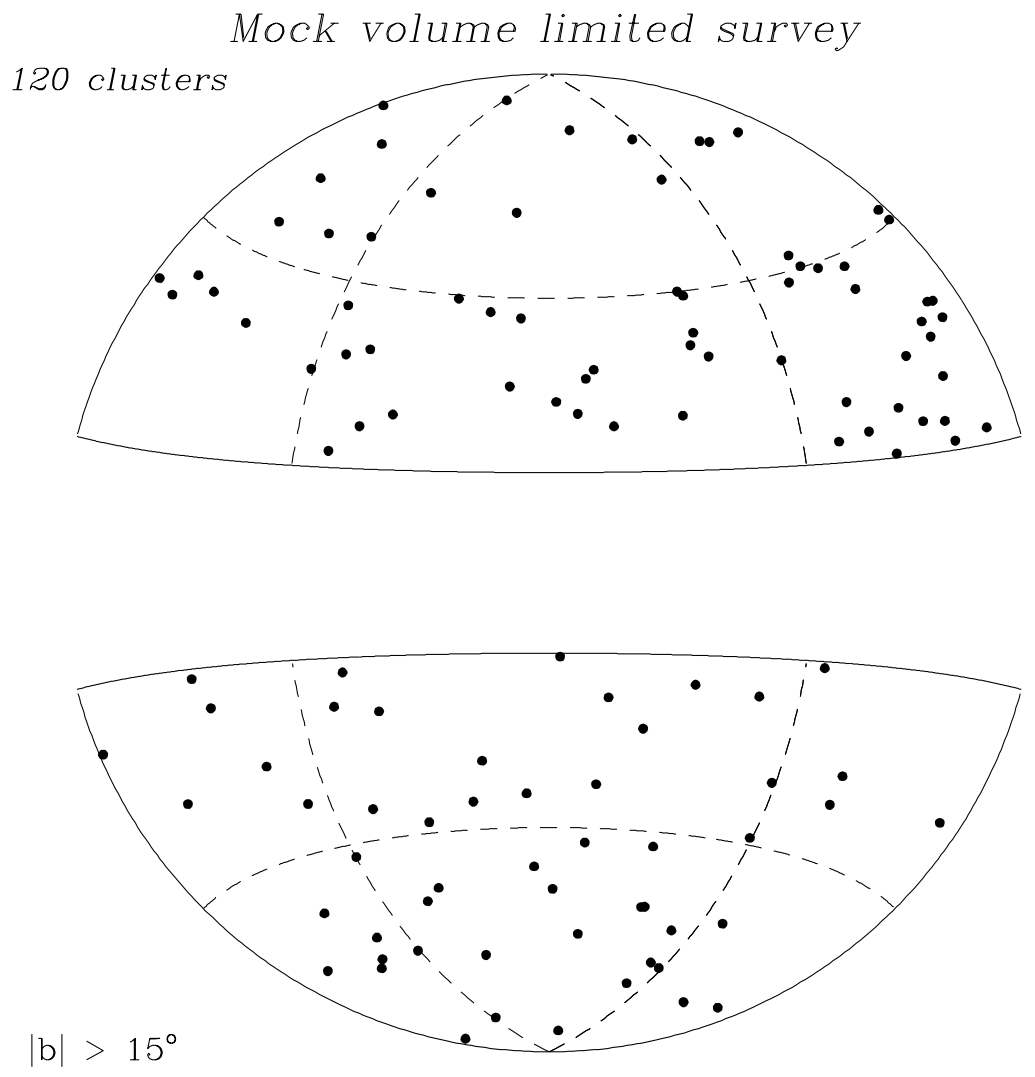

Fig. 1.— Aitoff–Hammer projection of a sample survey

For comparison with the LP survey, we construct a volume limited survey by placing $N = 120$ clusters at random within a sphere of radius $R = 150\mathrm{h}^{-1}\mathrm{Mpc}$. (Below we will study how our results depend on the value of $R$.) As in the actual survey, we eliminate any cluster with galactic latitude $< 15°$. A typical "mock" survey constructed in this way is shown in Fig. 1.

The observed line-of-sight velocities for our sample clusters are assumed to have two sources of uncertainty; we assume that the peculiar velocities of the clusters are Gaussian distributed about the uniform streaming motion with 1−D dispersion $\sigma_*$ and that the observational errors are also Gaussian distributed and given by $\sigma_n$ for the $n$th cluster. We shall take $\sigma_n = \beta r$, so that the error in measurement is a fraction $\beta/H$ of the distance $r$ to the cluster, $\sigma_*$ to be a constant over the sample and $H = 100\mathrm{km/s/Mpc}$.

Given these assumptions, it can be shown [1] that the maximum likelihood solution for the uniform streaming motion $U_i$ of a cluster sample is given by

$$U_i = A_{ij}^{-1} \sum_s \frac{r_{s,j} S_s}{\sigma_s^2 + \sigma_*^2} \ , \tag{1}$$

where

$$A_{ij} = \sum_s \frac{\hat{r}_{s,i} \; \hat{r}_{s,j}}{\sigma_s^2 + \sigma_*^2} \; . \qquad (2)$$

Here the cluster labeled by an index $s$ has position $\vec{r}_s$ and estimated line-of-sight peculiar velocity $S_s$ that is related to the true velocity by

$$S_s = \hat{r}_{s,i} \; v_i(\vec{r}_s) + \varepsilon_s \; , \qquad (3)$$

where $\varepsilon_s$ is drawn from a Guassian with zero mean and variance $\sigma_\varepsilon^2 = \sigma_s^2 + \sigma_*^2$.

Writing the uniform bulk velocity in terms of Eqs. (1), (2) and (3) we get

$$U_i = U_i^{(v)} + U_i^{(\varepsilon)} = A_{ij}^{-1} \sum_s \frac{\hat{r}_{j,s} \; \hat{r}_{k,s}}{\sigma_s^2 + \sigma_*^2} v_k(\hat{r}_s) + A_{ij}^{-1} \sum_s \frac{\hat{r}_{j,s} \; \varepsilon_s}{\sigma_s^2 + \sigma_*^2} \; . \qquad (4)$$

Since the two terms are statistically independent, we can write the covariance matrix as

$$R_{ij} \equiv < U_i U_j > = R_{ij}^{(v)} + R_{ij}^{(\varepsilon)} \; . \qquad (5)$$

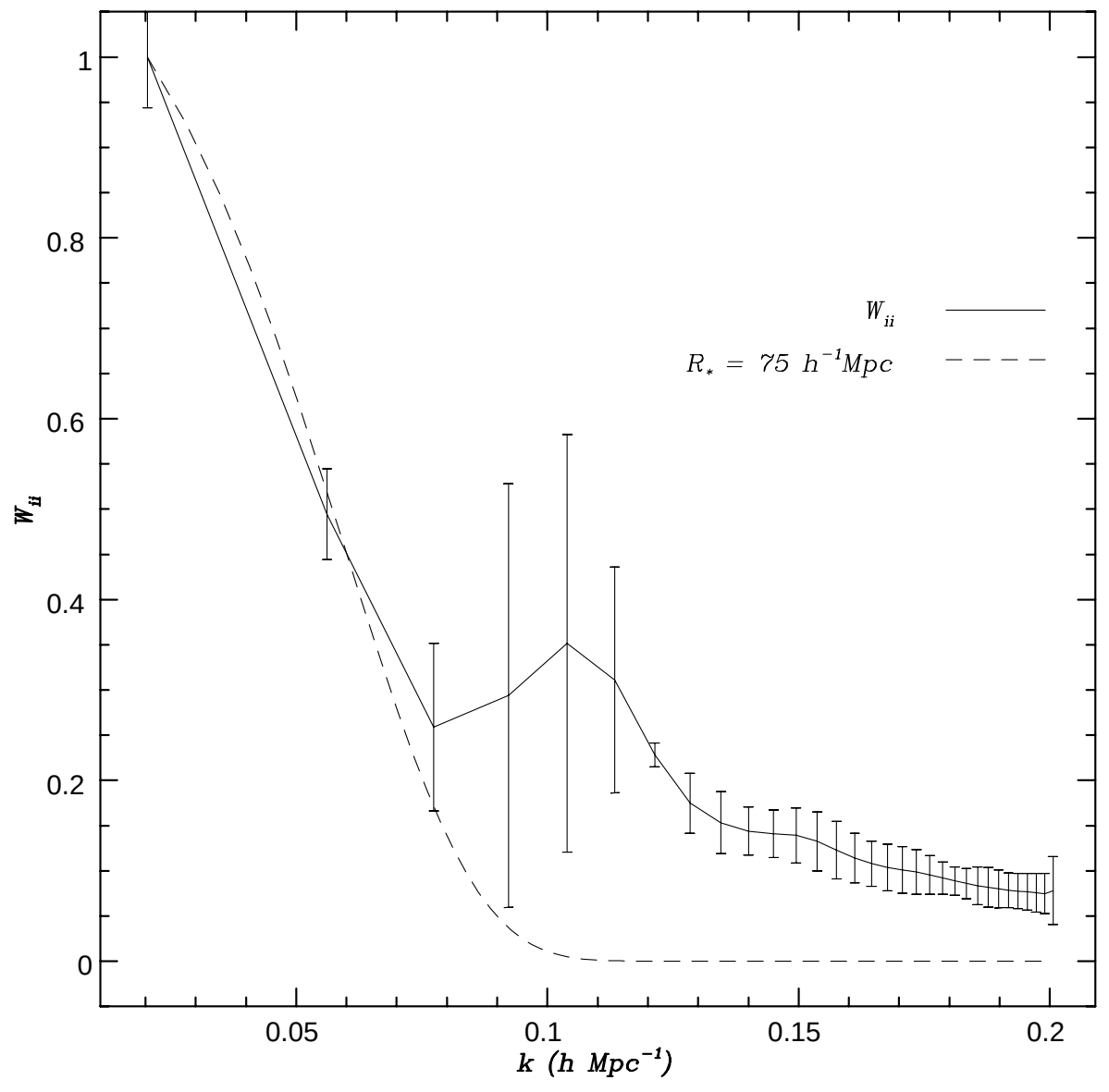

Fig. 2.— normalized diagonal elements of the squared tensor window function for the survey and a Gaussian $\propto e^{-k^2 R_*^2}$

The velocity term can be written as

$$U_i^{(v)}(\vec{r}_o) = \int d^3r \; W_{ij}(\vec{r}) \; v_j(\vec{r}_o + \vec{r}) \; , \qquad (6)$$

where $W_{ij}$ is the tensor window function for the survey, given in fourier space by

$$W_{ij}(\vec{k}) = A_{im}^{-1} \; \sum_s \frac{\hat{r}_{s,m} \; \hat{r}_{s,j}}{\sigma_s^2 + \sigma_*^2} \;\; e^{i\vec{k}\cdot\vec{r}} \; . \qquad (7)$$

Then for

$$\mathcal{W}_{ij}^2(\vec{k}) = W_{im}(\vec{k}) \; W_{jn}(\vec{k}) \; \hat{k}_m \; \hat{k}_n \; , \qquad (8)$$

the velocity part of the covariance matrix is a convolution of Eq. (8) and the velocity power spectrum

$$R_{ij}^{(v)} = \int d^3k \, \mathcal{W}_{ij}^2(\vec{k}) \, P_v(\vec{k}) \, , \tag{9}$$

where the velocity power spectrum is

$$P_v(k) \equiv < |v(\vec{k})|^2 > = \frac{H^2 a^2}{k^2} P(k) \, . \tag{10}$$

We use the density power spectrum $P(k)$ from the IRAS–QDOT survey [4] and its $1\sigma$ result; the unbiased CDM transfer function (BBKS) [5] ($\Omega h = 0.5$, $\sigma_8 = 1$); and the MDM simulations [6] normalized to the COBE quadrupole $Q_2 = 17\mu$K which is equivalent to a biasing parameter of $b = 1.5$.

In Fig. 2, we plot the window function calculated for an LP type sample constructed as described above. The window function is fit well by a Gaussian with $R_* \approx 75\mathrm{h}^{-1}$Mpc. In Fig. 3 we plot same window function superposed with power spectra from models of interest.

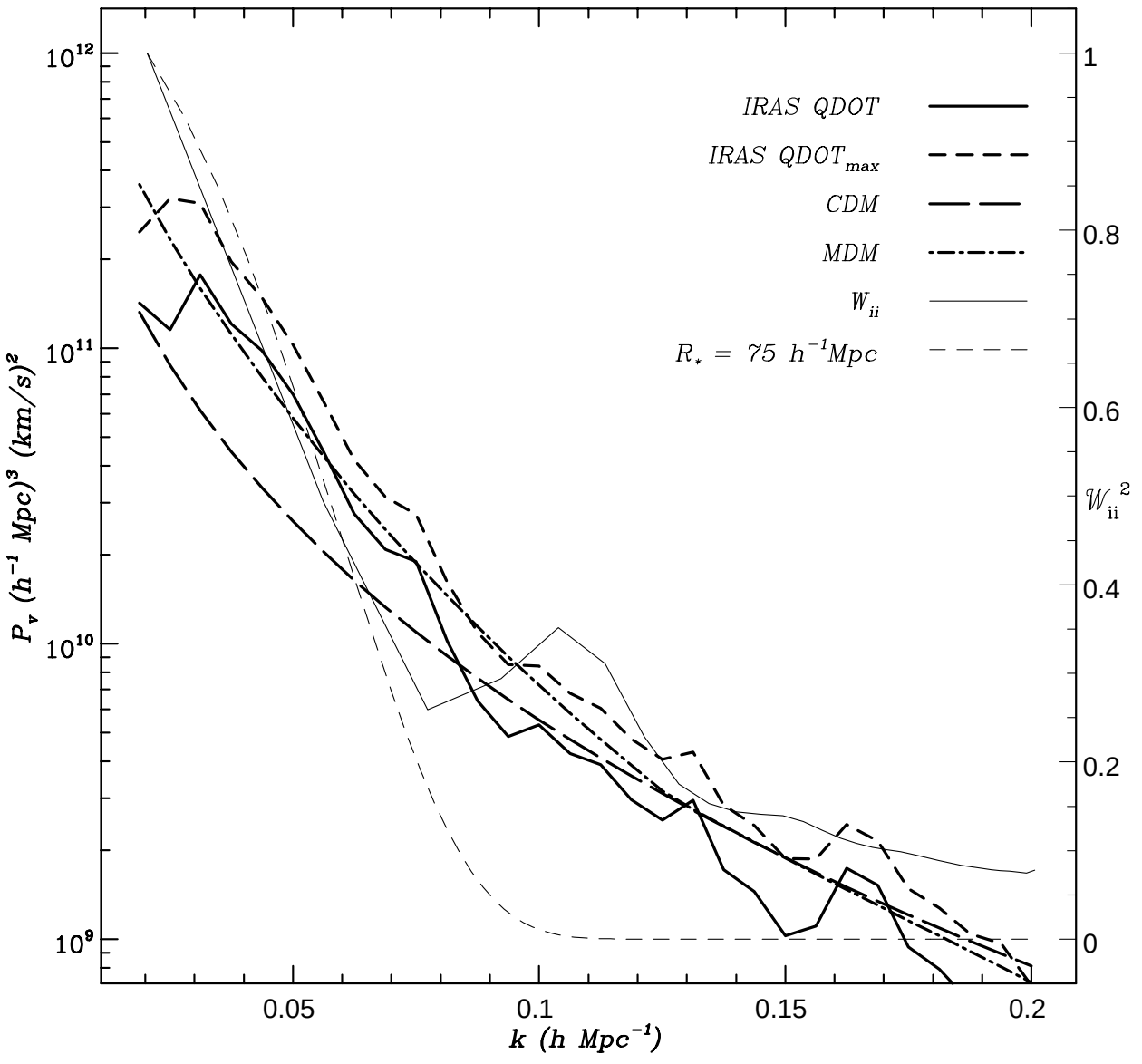

Fig. 3.— Velocity power spectra, the window function for the survey and a Gaussian $\propto e^{-k^2 R_*^2}$

The noise term is

$$R_{ij}^{(\varepsilon)} = \left\langle \left( A_{il}^{-1} \sum_s \frac{\hat{r}_{s,l} \ \hat{r}_{s,i}}{\sigma_s^2 + \sigma_*^2} \right) \left( A_{jm}^{-1} \sum_{s'} \frac{\hat{r}_{s',m} \ \hat{r}_{s',j}}{\sigma_{s'}^2 + \sigma_*^2} \right) \right\rangle = A_{ij}^{-1} \, . \tag{11}$$

Thus the final covariance matrix is

$$R_{ij} = \int d^3k \, \mathcal{W}_{ij}^2(\vec{k}) \, P_v(\vec{k}) + A_{ij}^{-1} \tag{12}$$

The quantity

$$\Lambda \equiv \sqrt{R_{ii}} \, ; \quad \Lambda^{(v)} = \sqrt{R_{ii}^{(v)}} \, , \tag{13}$$

gives the variance of the distribution for the magnitudes of uniform streaming velocities for a given sample. Given $\Lambda$ one can calculate the probability for observing a streaming velocity of a given magnitude for a particular survey type and power spectrum.

## 3. Large $N$ Limit

In the limit of large $N$, we can replace many of the sums in section 2 with integrals. This gives us an alternative method of calculating the covariance matrix which acts as a check on the analysis described above. In what follows, we shall assume that we have $N \gg 1$ clusters distributed randomly in a sphere of radius $R$. For simplicity we shall concentrate on the two cases $\sigma_* = 0$ and $\sigma_s = 0$. For the first case we shall take $\sigma_s = \beta r$.

To lowest order in $1/N$, we can write Eq. (2) as

$$A_{ij} = \sum_s \frac{\hat{r}_{s,i}\ \hat{r}_{s,j}}{\sigma_s^2 + \sigma_*^2} = \delta_{ij}\ \frac{N}{4\pi R^3}\int_0^R d^3r\ \frac{1}{\beta^2 r^2 + \sigma_*^2}\ +\ O(\sqrt{N})\ . \tag{14}$$

where we have used the fact that in the large $N$ limit, $A$ should be isotropic so that $A_{ij} = \delta_{ij}\ \sum_n A_{nn}/3$. For our two cases of interest

$$A_{ij} \approx \begin{cases} \delta_{ij}\,(N/R^2\beta^2) & \text{for } \sigma_* = 0 \\ \delta_{ij}\,(N/3\sigma_*^2) & \text{for } \sigma_s = \beta = 0 \end{cases}\ . \tag{15}$$

Next we will calculate $W_{ij}(\vec{k})$ using this method. Here we shall assume that the power spectrum $P_v(k)$ is a function only of the magnitude of $\vec{k}$. With this assumption, we need only calculate the angle averaged $W_{ij}(\vec{k})$. In what follows we shall take all quantities to be averaged over angles and to be functions only of the magnitude of $\vec{k}$. In this approximation $W_{ij}(k) = \delta_{ij}\ \sum_n W_{nn}(k)/3$. We then write Eq. (7) as

$$W_{ij}(k) = A^{-1}_{im}\ \delta_{mj}\ \frac{N}{4\pi}\int_0^R d^3r\ \frac{\exp\left(ikr\cos(\theta)\right)}{\beta^2 r^2 + \sigma_*^2}\ +\ O(\sqrt{N})\ , \tag{16}$$

where we have taken $\vec{k} = k\hat{z}$ for simplicity of calculation.

In our angle averaged approximation, we make the replacement $\hat{k}_m\hat{k}_n = \delta_{mn}/3$ in Eq. (8). Performing the integrals and using the expressions for $A_{ij}$ derived above we then find

$$\mathcal{W}^2_{ij}(k) \approx \begin{cases} \delta_{ij}\ 3\left(\sin(kR) - kR\cos(kR)\right)^2/(kR)^6 & \text{for } \sigma_* = 0 \\ \delta_{ij}\ \left(\mathrm{Si}(kR)\right)^2/(kR)^2 & \text{for } \sigma_s = 0 \end{cases} \tag{17}$$

where $\mathrm{Si}(x) = \int_0^x dy\ \sin(y)/y$.

Given a power spectrum, we can use the $A_{ij}$ and $\mathcal{W}^2_{ij}(k)$ derived above to calculate $R_{ij}$ and thus $\Lambda$ in the large $N$ limit.

In realistic surveys, full sky coverage is never achieved. For the LP survey, for example, a $15^\circ$ wedge about the galactic disk is removed, thus covering only $\approx 75\%$ of the sky. For the noise term, $A^{-1}_{ij}$, we can account for this by multiplying our result by $1/sqrt(0.75)$. For the other term, things are somewhat more complicated. We shall explore this question more thoroughly in an upcoming paper [7].

## 4. Results

We have applied the formalism of section 2 to a set of "mock" surveys selected to mimic the features of the LP survey. For a given mock survey and power spectrum, the variance of the streaming velocity $\Lambda$ can be calculated as described above. Due to statistical fluctuations, the value of $\Lambda$ thus calculated will vary for different random realizations of a survey. From §3 above we expect the variation in $\Lambda$ over different realizations to be of order $\Lambda/\sqrt{N} \sim 0.1\Lambda$.

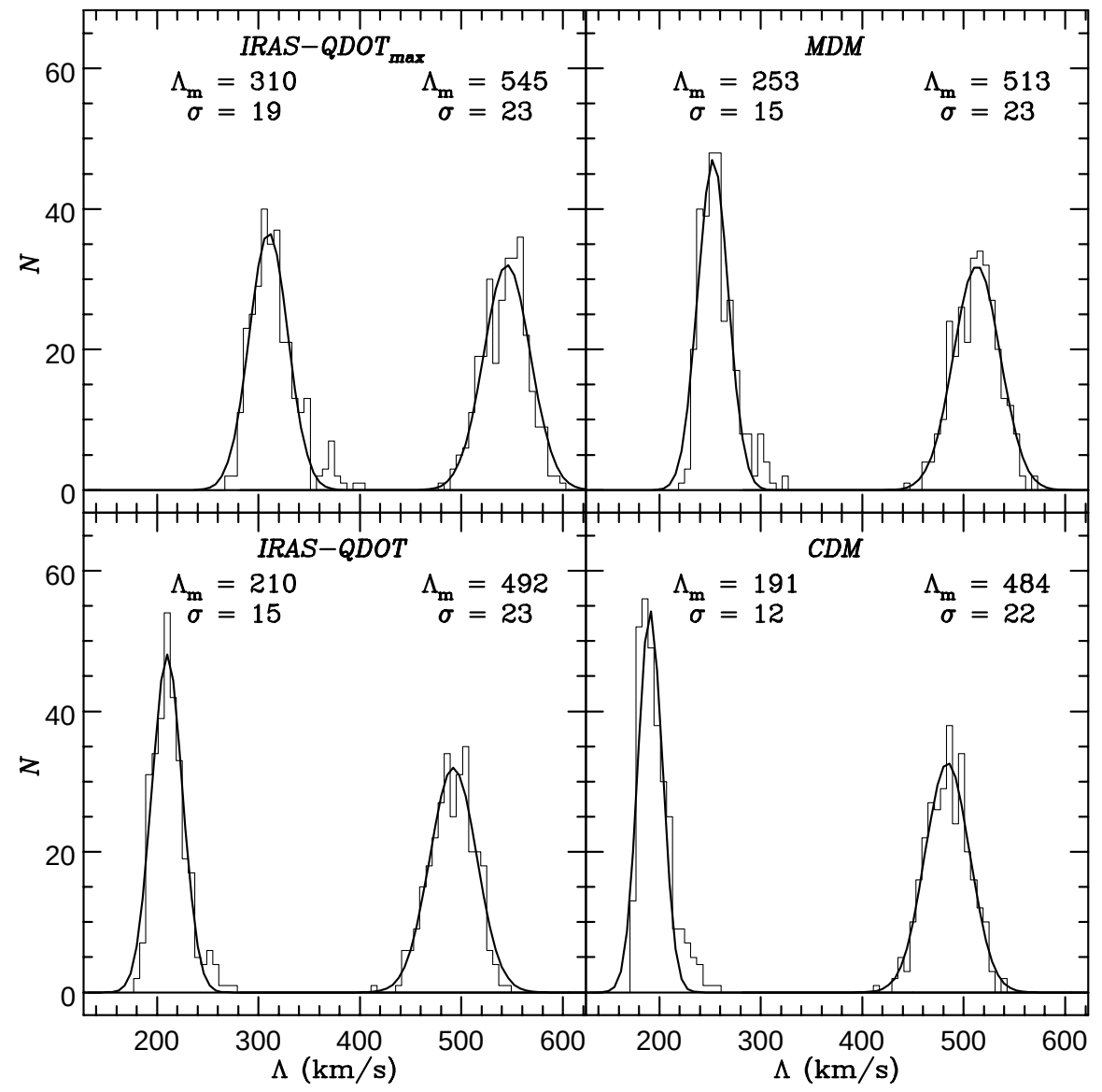

Fig. 4.— The distribution of $\Lambda$ for the four power spectra we used for both $\Lambda^{(v)}$ (the true variance) and $\Lambda$ (the left and right distribution respectively). The parameters for the Gaussian fit are shown.

In Fig. 4 we show the result of 300 realizations of a LP type survey, with $\beta = 16$km/s/Mpc, $\sigma_* = 300$km/s, radius $R = 15000$km/s and $N = 120$ clusters. The distribution of $\Lambda$ over the different realizations is well fit by a Gaussian $e^{-(\Lambda-\Lambda_m)^2/2\sigma^2}$ and in good agreement with expectations. In Fig. 5 we show the variances as a function of the radius of the survey ball and the percent uncertainty $\beta$ for the four velocity power spectra. As expected, the true variance (*i.e.* $\beta$=0), indeed decreases as the size of the frame increases. However, for more realistic surveys, where the effect of the noise is hard to estimate accurately, the variance increases at large radii.

In Fig. 6 we compare the results from the numerical analysis to the analytical results presented in §3 for the CDM[5] power spectrum with $\Omega h = 0.5$ and $\sigma_8 = 1$. As can be seen, the numerical and analytical analyses give similar results. The discrepancy for small $R$ arises because in the analytical analysis we set $\sigma_* = 0$ for $\sigma_s \neq 0$ which has an affect in the small $R$ regime.

In the table below we give the results for the measured variance $\Lambda$ and the noise–free variance $\Lambda^{(v)}$. Next to them we present the likelihood of getting the LP velocity of 730km/s and the chance of getting this result, *i.e.* 100− CL (confidence level) of the result. The chances of finding large amplitude velocities in a survey of the LP type are quite reasonable $> 10\%$ (*i.e.* these are $1.3 - 1.5\sigma$ results) given the power spectra we used. However, the chance of these velocities to be the actual, true velocity for the frame are quite small $< 2\%$ ($2.3 - 3.9\sigma$ result).

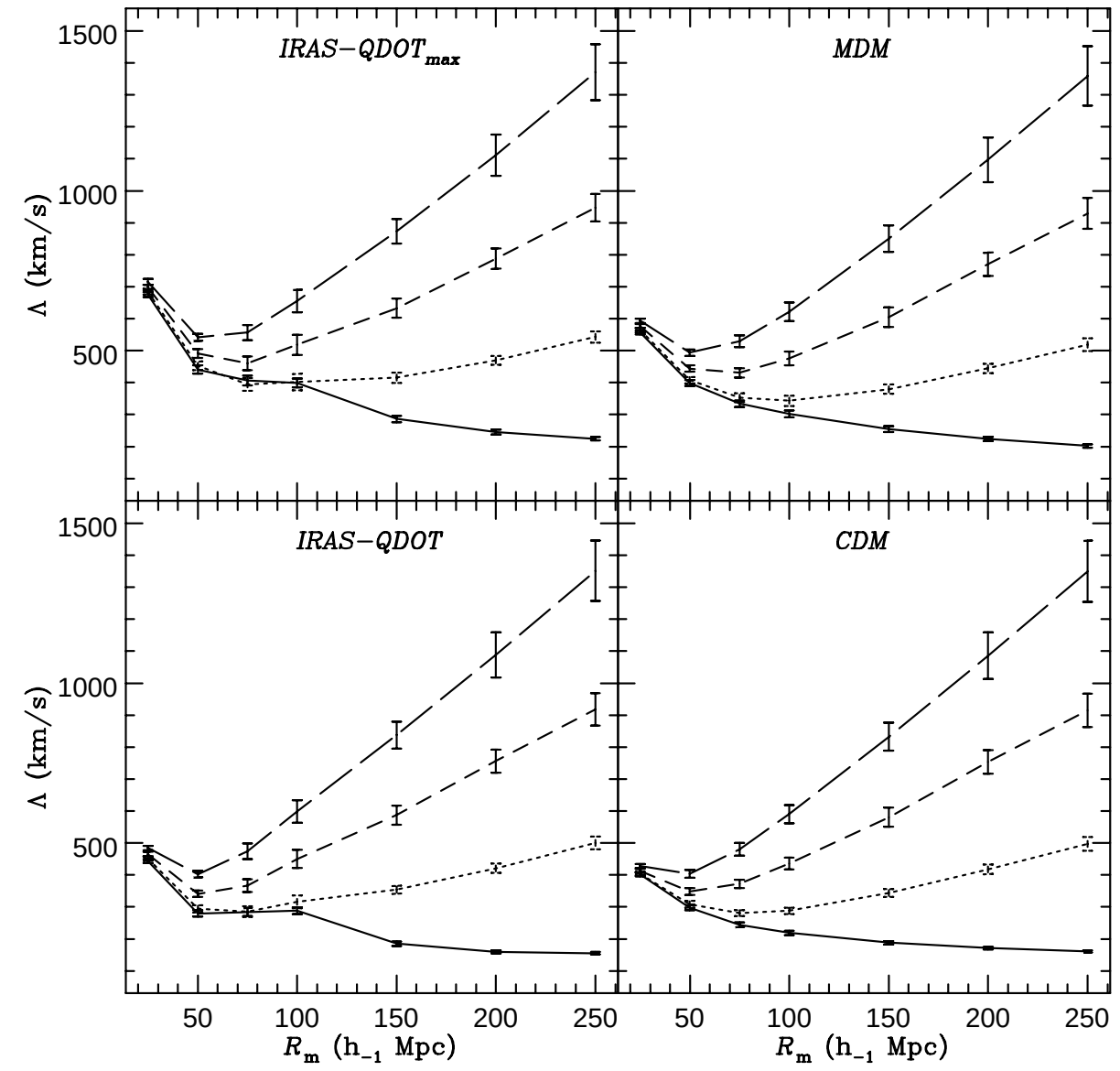

Fig. 5.— The expected values of $\Lambda$ for the four power spectra we used as a function of the maximum radius of the survey and the percent error in the the velocity dispersion. The solid curve is the true variance $\Lambda^{(v)}$, the dotted, short dashed and long dashed lines are for $\beta/H =$ 10%, 20% and 30% respectively.

| Spectrum | $\Lambda$ (km/s) | $\sigma$ | % | $\Lambda^{(v)}$ (km/s) | $\sigma^{(v)}$ | % |
|---|---|---|---|---|---|---|
| IRAS | $492 \pm 23$ | $1.48^{+0.07}_{-0.07}$ | $13.79^{+1.85}_{-1.83}$ | $210 \pm 15$ | $3.48^{+0.23}_{-0.27}$ | $0.05^{+0.07}_{-0.03}$ |
| IRAS$_{\rm max}$ | $545 \pm 23$ | $1.34^{+0.05}_{-0.06}$ | $18.04^{+1.83}_{-1.85}$ | $310 \pm 19$ | $2.35^{+0.14}_{-0.15}$ | $1.85^{+0.80}_{-0.64}$ |
| CDM | $484 \pm 22$ | $1.51^{+0.06}_{-0.07}$ | $13.15^{+1.76}_{-1.74}$ | $191 \pm 12$ | $3.82^{+0.23}_{-0.26}$ | $0.01^{+0.02}_{-0.01}$ |
| MDM | $513 \pm 23$ | $1.42^{+0.06}_{-0.07}$ | $15.47^{+1.85}_{-1.85}$ | $253 \pm 15$ | $2.88^{+0.16}_{-0.18}$ | $0.39^{+0.25}_{-0.17}$ |

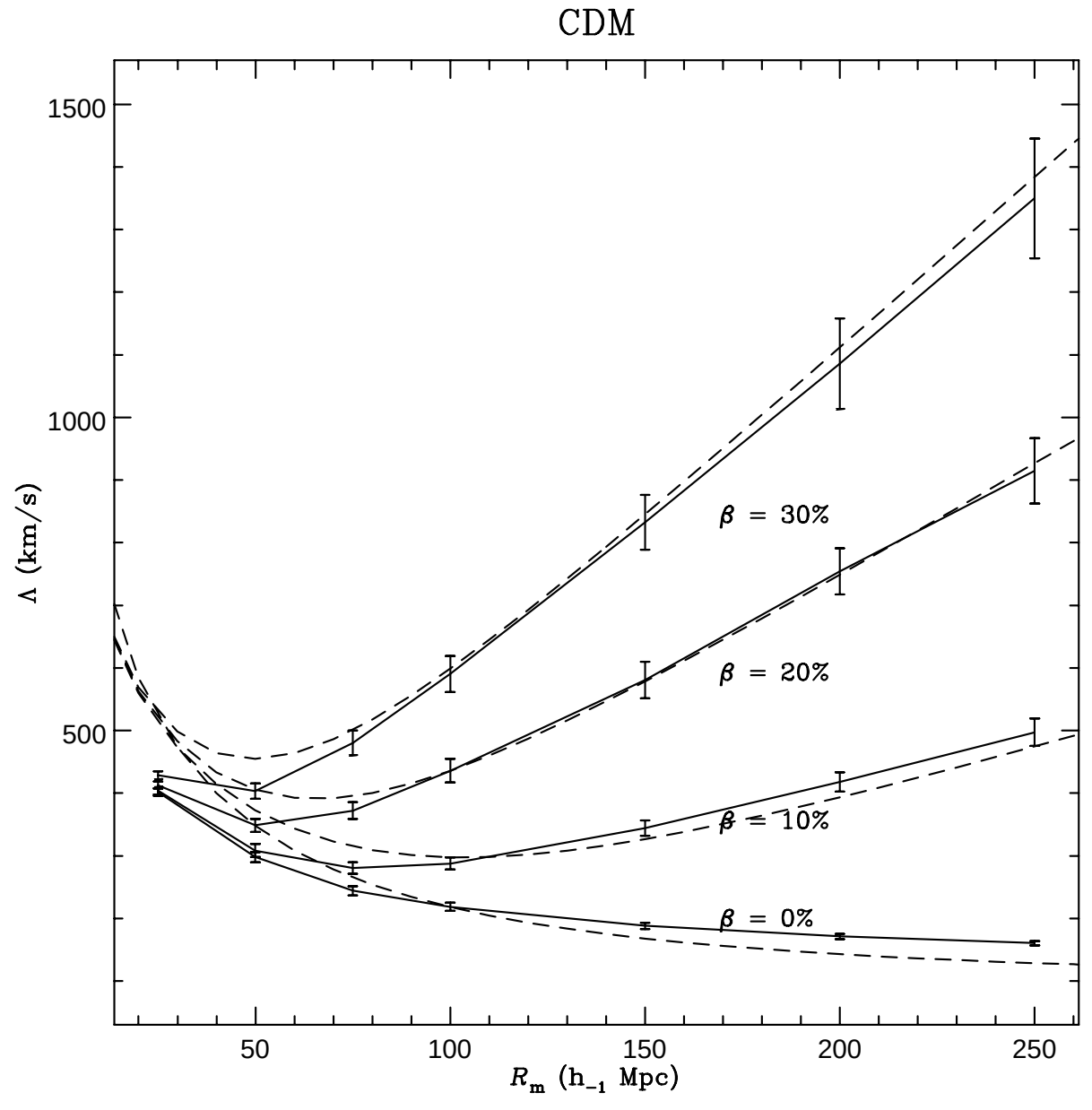

Fig. 6.— The numerical and analytical results of $\Lambda$ for the CDM power spectrum ($\Omega h = 0.5 \quad \sigma_8 = 1$) used as a function of the maximum radius of the survey and the percent error in the the velocity dispersion $\beta$. The solid curves are the numerical reults, the dashed line are the analytical ones.

## 5. Conclusions

We have presented a formalism to calculate the velocity variance as a function of the geometry of the survey. We show that the variance is a quadrature superposition of the true velocity and the noise inherent in the observation. We apply the formalism to realizations of surveys similar to the Lauer & Postman analysis [2]. We show that comparing there results to the expected variance for this type of survey we find that there velocity is a $1-2\sigma$ result depending on the power spectra we used, whereas comparing it to true expected velocity we get a $2.3-3.9\sigma$ result. It is indeed true that we do not know the power spectrum for this type of clusters and thus are unable to make a definitive statement about the expected velocities in this kind of frame. However, we believe that it is not unreasonable to assume that the power spectra we used are quite good estimates, and the true power is not significantly larger on scales $> 75\mathrm{h}^{-1}$Mpc than the $1\sigma$ IRAS–QDOT one. We therefore conclude that although it is not unlikely that LP would measure a velocity of the magnitude they quote without the need for extra power on large scales, the true velocity of the frame relative to the cosmic rest frame is likely to be significantly lower. Clearly, if the LP results are confirmed and represent the actual bulk flow of the Abell frame then the models presented here will be ruled out at high confidence level; further, any susccessful model will have to address the discrepency between the IRAS–QDOT and LP results. We will present a more rigorous analysis of the problem and include the clustering properties of the Abell catalogue in a subsequent paper [7].

**Acknowlegement:** We thank Nick Kaiser for valuable discussions and helpful conversations. HAF was supported in part by the National Science Foundation grant NSF–PHY–92–96020. RW was supported in part by NASA grant NAGW–2802.

---